\documentclass[aps,prd,twocolumn,showpacs,preprintnumbers,amsmath,amssymb,showpacs,showkeys]{revtex4}
\usepackage{graphicx}% Include figure files
\usepackage{dcolumn}% Align table columns on decimal point
\usepackage{bm}% bold math
\usepackage{amssymb}
\usepackage{indentfirst}
\usepackage{epsfig}
\usepackage{epsf}
\usepackage{graphicx}
\usepackage{slashbox}

\newcommand{\eq}{\begin{eqnarray}}
\newcommand{\en}{\end{eqnarray}}
\newcommand{\bfb}{{\bf b}_{\perp}}
\newcommand{\bfk}{{\bf k}_{\perp}}
\newcommand{\bfbj}{{\bf b}_{\perp j}}

\newcommand{\ra}{\rangle}

\begin{document}

\title{Meson wave function from holographic models}

\author{Alfredo Vega$^{1}$,
        Ivan Schmidt$^{1}$,
        Tanja Branz$^{2}$,
        Thomas Gutsche$^{2}$,
        Valery E. Lyubovitskij$^{2}$\footnote{On leave of absence
from Department of Physics, Tomsk State University, 634050 Tomsk, Russia}
\vspace*{1.2\baselineskip}}

\affiliation{$^{1}$Departamento de F\'\i sica y Centro de Estudios
Subat\'omicos,  Universidad T\'ecnica Federico Santa Mar\'\i a,
Casilla 110-V, Valpara\'\i so, Chile \\
\vspace*{.2\baselineskip} \\
$^{2}$ Institut f\"ur Theoretische Physik,
Universit\"at T\"ubingen,\\
Kepler Center for Astro and Particle Physics, \\
Auf der Morgenstelle 14, D--72076 T\"ubingen, Germany\\}

\date{\today}

\begin{abstract}

We consider the light-front wave function for the valence quark state
of mesons using the AdS/CFT correspondence, as has been suggested by
Brodsky and T\'eramond. Two kinds of wave functions, obtained in different
holographic Soft-Wall models, are discussed.

\end{abstract}

\preprint{USM-TH-247}

\maketitle

\section{Introduction}

The hadronic wave function in terms of quark and gluon degrees of
freedoms plays an important role in QCD process predictions.
For example, knowledge of the wave function allows to calculate
distribution amplitudes and structure functions or conversely these
processes can give phenomenological restrictions on the wave functions.

In principle the Bethe-Salpeter approach~\cite{BetheSalpeter} and
discrete quantization in the light-front formalism~\cite{PauliBrodsky}
allow to obtain hadronic wave functions but in practice several
problems present to realize this~\cite{BJS,JK}. Therefore approximate
solutions for hadronic bound states are usually considered using in a first
step specific quarks models to obtain the valence quark wave function.

There are several non-perturbative approaches to ob\-ta\-in properties of
distribution amplitudes and/or hadronic wave functions from QCD, and now
we have possibility to include techniques based on the
Anti-de Sitter space/conformal field theory (AdS/CFT) correspondence.

Although a rigorous QCD dual is unknown, a simple approach known as
Bottom-Up allows to built models that have some essential QCD features,
including counting rules at short and confinement at long distances.
This model has been successful in several QCD applications such as
hadronic scattering processes~\cite{PolStrass1,Janik,BdT1,Levin},
hadron spectrum~\cite{BdT2,KKSS,Forkel,VegaSchmidt1,VegaSchmidt2},
hadronic couplings and chiral symmetry breaking~\cite{DaRol,EKSS,Colangelo},
quark potentials~\cite{Boschi,Andreev,Jugeau} and
hadron decays~\cite{Hambye}.

Together with these applications it is possible to set up a mapping
between specific properties of the AdS description for hadrons and
the Hamiltonian formulation for quantized QCD in the light-front
formalism. Latter approch allows to obtain an excellent first approximation
to the valence wave function for mesons~\cite{BdT3, BdT4}. Wave functions
obtained using the AdS/CFT correspondence can be used as an initial ansatz
for a variational treatment or as basis states to diagonalize
the light-front QCD Hamiltonian.

In this work meson wave functions obtained in the context of AdS/CFT
ideas~\cite{BdT3,BdT4} are studied considering two kinds of holographic
Soft-Wall models. First we consider the more usual model with a
quadratic dilaton~\cite{KKSS,DaRol,BdT4}. Then we discuss
predictions of a recent model which considers a logarithmic dilaton
as suggested by Einstein's equations for an AdS metric. It also
includes anomalous dimensions~\cite{VegaSchmidt2} and allows to
reproduce the Regge behavior even in the baryonic sector.

The work is structured as follows. Sec.~II is devoted to the
extraction of wave functions for scalar/pseudoscalar mesons using the
two holographic models. In Sec.~III we concentrate on the pion wave
function discussing the adjustment of the model parameters. Distribution
amplitudes and parton distributions for the valence state are
calculated in both models. In the pion case we consider both current and
constituent quark masses. In Sec.~IV we calculate decay constants in
the simplified case when the valence component is dominant.
Conclusions are presented in Sec.~V.

\section{Meson wave function in holographical models}

The comparison of form factors calculated both in the light-front
formalism and in AdS offers the possibility to relate AdS modes to
light-front wave functions (LFWF)~\cite{BdT3,BdT4}. Below we briefly
discuss the derivation of this matching procedure.

\begin{figure*}
  \begin{tabular}{cc cc}
    \includegraphics[width=3.0in]{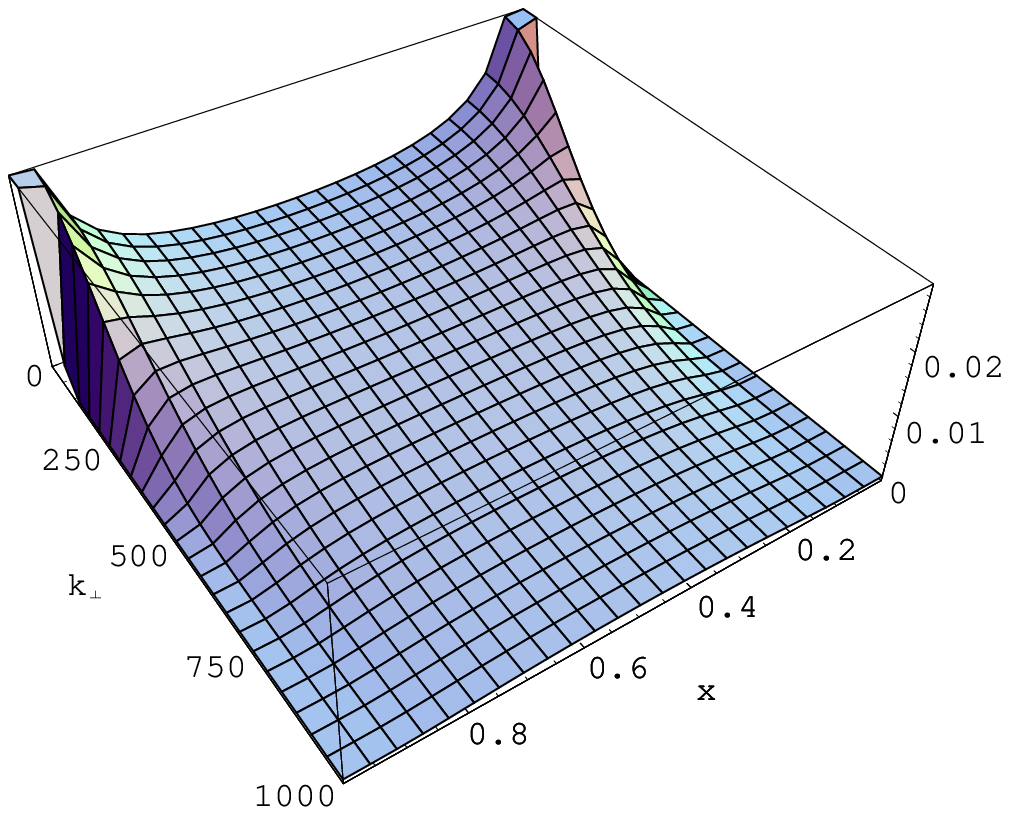} &
    \includegraphics[width=3.0in]{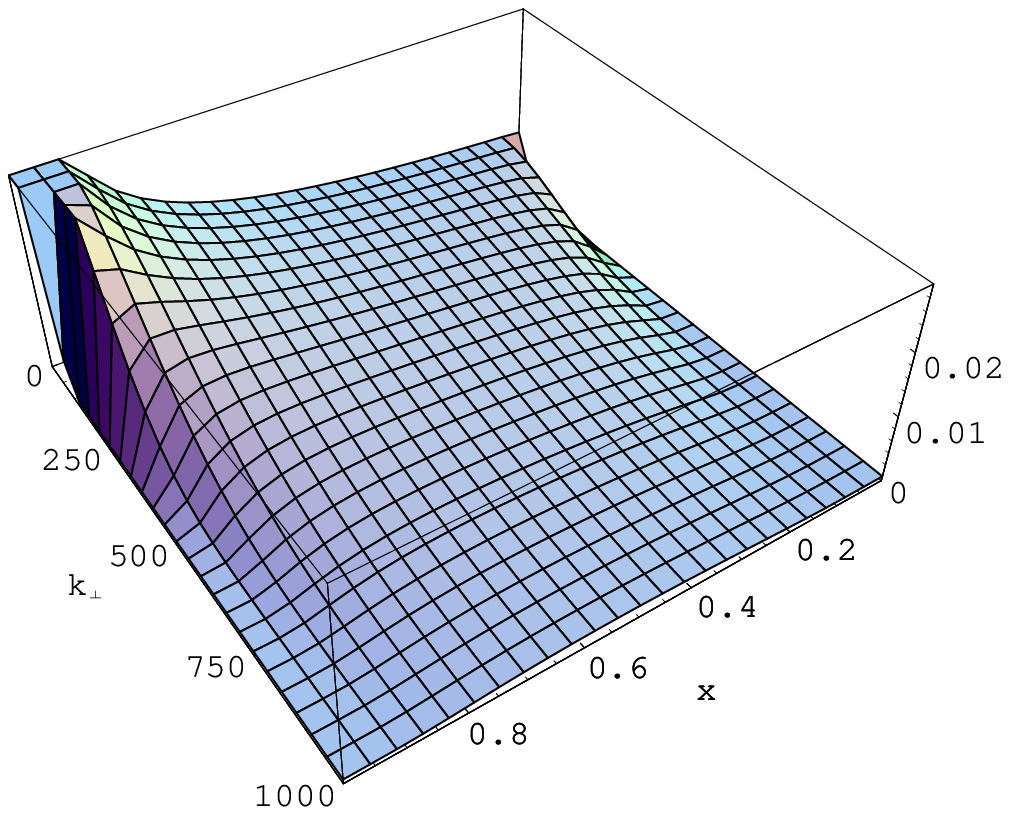}
  \end{tabular}
\caption{The pion wave function $\psi(x,\bfk)$, for $m = 4$ MeV.
The left graph corresponds to Eq.~(\ref{FnOnda1}) and the right
one to Eq.~(\ref{FnOnda2}). }
\end{figure*}

In the light-front formalism the electromagnetic form factor of pion
can be written as~\cite{BdT4}

\begin{equation}
 \label{FactorFormaLF}
 F(Q^{2}) = 2 \pi \int\limits^{1}_{0} dx \frac{1-x}{x}
\int\limits_0^\infty d\zeta \zeta J_{0}
\biggl(\zeta Q \sqrt{\frac{1-x}{x}}\biggr)
\widetilde{\rho}  (x, \zeta),
\end{equation}
where $\widetilde{\rho}(x, \zeta)$ is the effective transverse
distribution of partons; $Q^2$ is the spacelike momentum transfer
squared; $J_0$ is the Bessel function.
Here we introduced the variable
\begin{equation}
 \zeta = \sqrt{\frac{x}{1-x}}
\bigg| \sum^{n-1}_{j=1} x_{j} \bfbj \bigg| \,,
\end{equation}
which represents the $x$-weighted transverse impact coordinate
of the spectator system.

On the other side the corresponding expression for scalars in AdS
with a dilaton $\varphi(z)$ is
\begin{equation}
 \label{FactorFormaAdS}
 F(Q^2) = \int\limits_0^\infty dz \Phi(z) J_\kappa(Q^2, z) \Phi(z) \,,
\end{equation}
where $\Phi(z)$ corresponds to modes that represent hadrons,
$J(Q^2, z)$ is the dual mode to the electromagnetic current, and
the metric considered is
\begin{equation}\label{AdS_metric}
 ds^2 =  \frac{R^{2}}{z^{2}} \eta_{\mu\nu} dx^{\mu} dx^{\nu}\,,
\ \ \eta_{\mu\nu} = {\rm diag} (1,-1,-1,-1-1) \,,
\end{equation}
where $z$ is the holographic coordinate and $\kappa$ is
the scale parameter characterizing the dilaton field.
An important step is to set up the electromagnetic current as
\begin{equation}
J(Q^2, z) = \int^{1}_{0} dx f(x)
J_{0}\biggl(\zeta Q \sqrt{\frac{1-x}{x}}\biggr) \,,
\end{equation}
Putting $z = \zeta$ and comparing Eqs.~(\ref{FactorFormaLF}) and
(\ref{FactorFormaAdS}) we get
\begin{equation}
 \label{Densidad}
 \widetilde{\rho}(x, \zeta) =
\frac{x f(x)}{1-x} \, \frac{|\Phi (\zeta)|^{2}}{2\pi\zeta} \,.
\end{equation}
Finally, considering the case with two partons $q_1$ and $\bar q_2$
\begin{equation}
 \label{Densidad2Partones}
 \widetilde{\rho}_{n=2}(x, \zeta) =
\frac{| \widetilde{\psi}_{q_1\bar q_2}(x,\zeta) |^{2}}{(1-x)^{2}}
\, \frac{1}{A^2} \,,
\end{equation}
where $\zeta^{2} = x (1-x) \bfb^2$ and $A$ is the normalization constant,
we obtain the relation between the AdS modes and the
meson LFWF $\widetilde{\psi}_{q_1\bar q_2}(x,\zeta)$
\begin{equation}
 \label{MapeoFnOnda}
 | \widetilde{\psi}_{q_1\bar q_2}(x,\zeta) |^{2} = A^2 \,
 x (1-x) f(x) \frac{|\Phi(\zeta)|^{2}}{2\pi\zeta} \,.
\end{equation}
Here $A$ is constrained by the probability condition
\eq\label{Probability_cond}
P_{q_1\bar q_2} = \int\limits_0^1 dx \int d^2 \bfb
|\widetilde{\psi}_{q_1\bar q_2}(x,\bfb)|^2 \leq 1
\en
with $P_{q_1\bar q_2}$ being the
probability of finding the valence Fock state $|q_1 \bar q_2\ra$
in the meson $M$. Note, in the case of massless quarks we have
$A = \sqrt{P_{q_1\bar q_2}}$, while this is not the case for
massive quarks (see discussion in Sec.IIA).
Next we consider two kinds of holographical models (Model 1 and
Model 2) and their respective wave functions.
\begin{figure*}
    \begin{tabular}{cc cc}
    \includegraphics[width=3.0in]{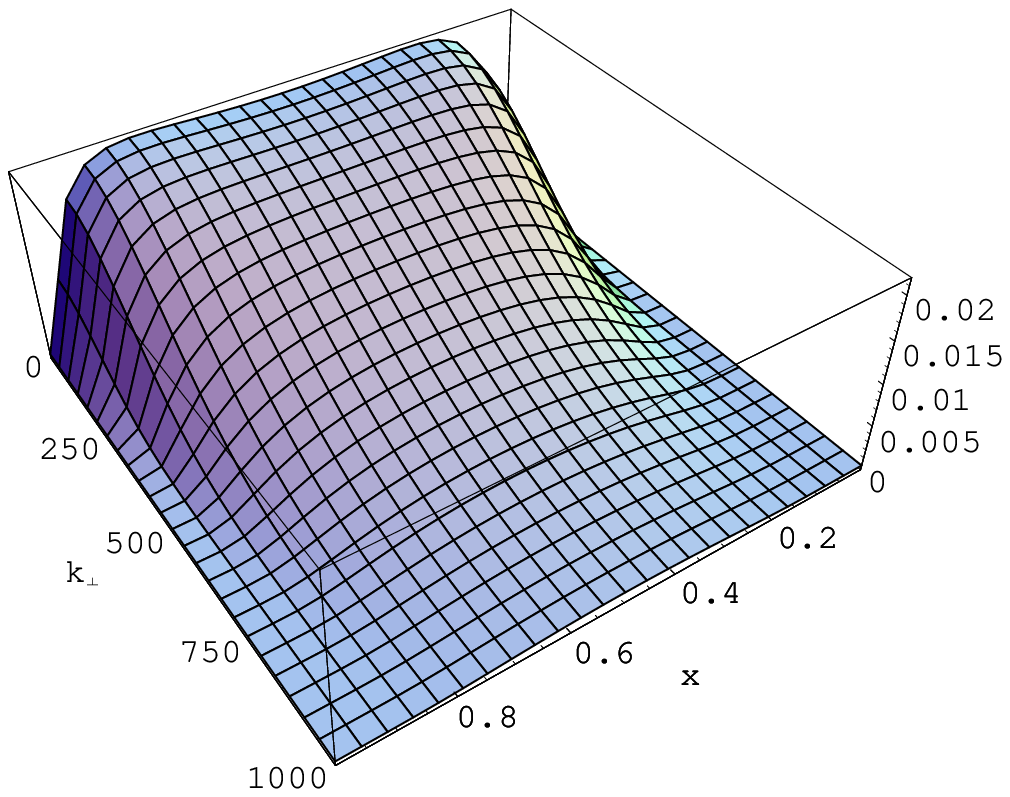}
  & \includegraphics[width=3.0in]{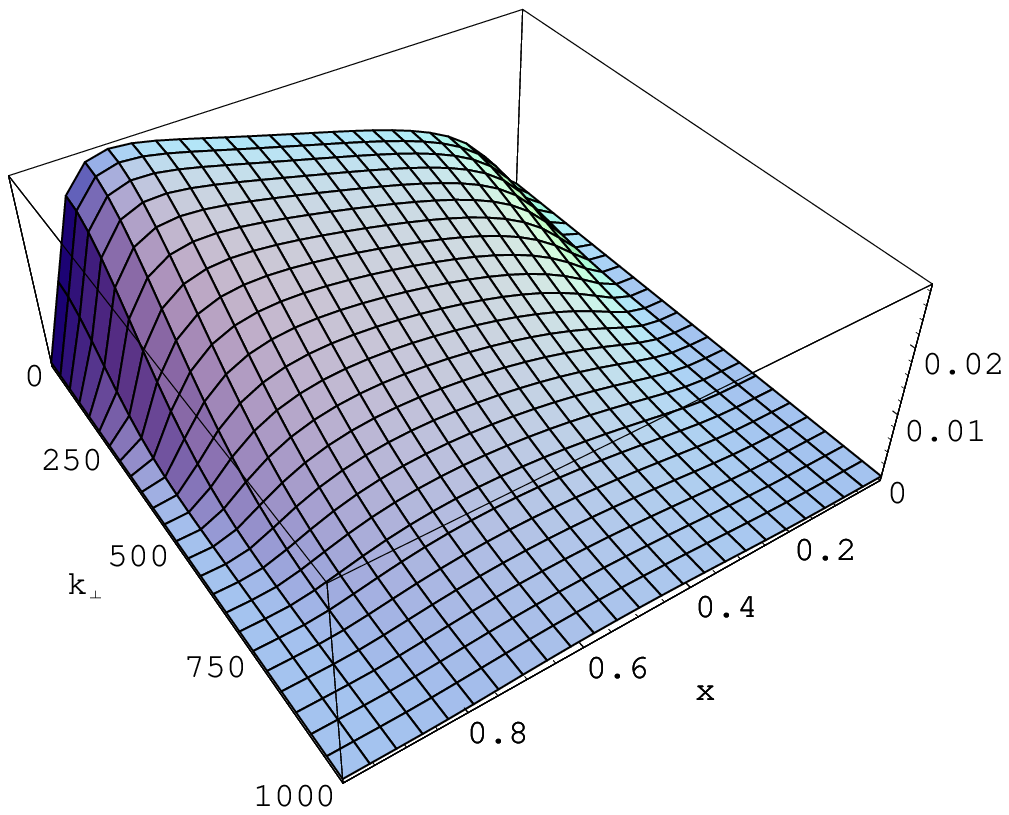}
    \end{tabular}
\caption{The pion wave function $\psi_{\pi}(x,\bfk)$,
for $m = 330$ MeV. The left graph corresponds to
equation Eq.~(\ref{FnOnda1}) and the right one
to Eq.~(\ref{FnOnda2}). }
\end{figure*}

\subsection{Model 1}

Model 1 is based on the Schr\"odinger equation~\cite{BdT6}
\eq
 \label{EcuationBdT}
 & &\biggl[ -\frac{d^{2}}{d \zeta^{2}} -
 \frac{1 - 4 L^{2}}{4 \zeta^{2}} + \kappa^4_1 \zeta^{2}
 + 2 \kappa^2_1 (L + S -1) \biggr] \Phi_1(\zeta) \nonumber\\
&=& M^2_1 \Phi_1(\zeta) ,
\en
for the AdS modes $\Phi(\zeta)$ that describe hadrons with integer
spin $S$ and the mass spectrum
\begin{equation}
 \label{EspectroBdT}
 M^2_1 = 4 \kappa^2_1 \biggl( n + L + \frac{S}{2} \biggr) \,,
\end{equation}
where $n$ and $L$ are the radial and orbital quantum numbers.
Here subscript ``1'' indicates the solutions of Model~1.

In this model the function $f(x)$ in matching condition~(\ref{MapeoFnOnda})
is fixed as $f(x)=1$ for large values of $Q^2 \gg 4 \kappa^2$. In this case
the current $J_\kappa(Q^2, z)$ decouples from the dilaton field \cite{BdT4}.
The examples considered in this work correspond
to mesons with $n = L = 0$, although both for scalars and vectors we find
\begin{equation}\label{Phi1_zeta}
\Phi_1(\zeta) = \kappa_1 \sqrt{2\zeta}
\exp^{-\frac{1}{2} \kappa^2_1 \zeta^2} \sim
\sqrt{\zeta} \exp^{-\frac{1}{2} \kappa^2_1 \zeta^2} \,.
\end{equation}
Using Eq.~(\ref{Phi1_zeta}) and keeping in mind that
$\zeta^2 = x (1-x) \bfb^2$, the meson LFWF of this model is
\begin{equation}
 \label{FnOndaBdTSinMasa}
 \widetilde{\psi}_{q_1\bar q_2}^{(1)}(x,\bfb)
= \frac{\kappa_1 A_1}{\sqrt{\pi}}
\sqrt{x(1-x)} \exp\Big(- \frac{1}{2} \kappa^2_1 x(1-x) \bfb^2\Big) \,.
\end{equation}
The wave function~(\ref{FnOndaBdTSinMasa}) does not consider
massive quarks. We include the quark masses following the prescription
suggested by Brodsky and T\'eramond~\cite{BdT5}. First one should perform
the Fourier transform of~(\ref{FnOndaBdTSinMasa})
\begin{equation}
\psi_{q_1\bar q_2}^{(1)}(x,\bfk) = \frac{4\pi A_1}{\kappa_1 \sqrt{x(1-x)}}
\exp\biggl(-\frac{\bfk^2}{2 \kappa^2_1 x(1-x)}\biggr) \,.
\end{equation}
In a second step the quark masses are introduced by extending the kinetic
energy of massless quarks with $K_0 = \frac{\bfk^2}{x(1-x)}$
to the case of massive quarks:
\eq\label{IncluyeMasas}
K_0 \to K = K_0 + \mu_{12}^2\,, \ \ \
\mu_{12}^2 = \frac{m^2_1}{x} + \frac{m^2_2}{1-x} \,.
\en
Note, the change proposed in~(\ref{IncluyeMasas}) is equivalent to
the following change in~(\ref{EcuationBdT})
\begin{equation}
- \frac{d^{2}}{d \zeta^{2}} \rightarrow
- \frac{d^{2}}{d \zeta^{2}} + \mu_{12}^2 \,.
\end{equation}
Finally we obtain
\begin{equation}
 \label{FnOnda1}
\hspace*{-.2cm}
 \psi_{q_1\bar q_2}^{(1)}(x,\bfk)
= \frac{4\pi A_1}{\kappa_1 \sqrt{x(1-x)}}
  \exp\biggl(-\frac{\bfk^2}{2 \kappa_1^2 x(1-x)}
- \frac{\mu_{12}^2}{2\kappa^2_1}
 \biggr) \,.
\end{equation}
Note, in the case of massive quarks the normalization constant fulfills
the relation
\eq
A_1 = \sqrt{P_{q_1\bar q_2}} \,
\Big(\int\limits_0^1 dx e^{-\mu_{12}^2/\kappa^2_1}\Big)^{-1/2}
\en
and $A_1 \to \sqrt{P_{q_1\bar q_2}}$ when $m_{1,2} \to 0$.

\begin{figure*}
  \begin{tabular}{cc cc}
    \includegraphics[width=3.0in]{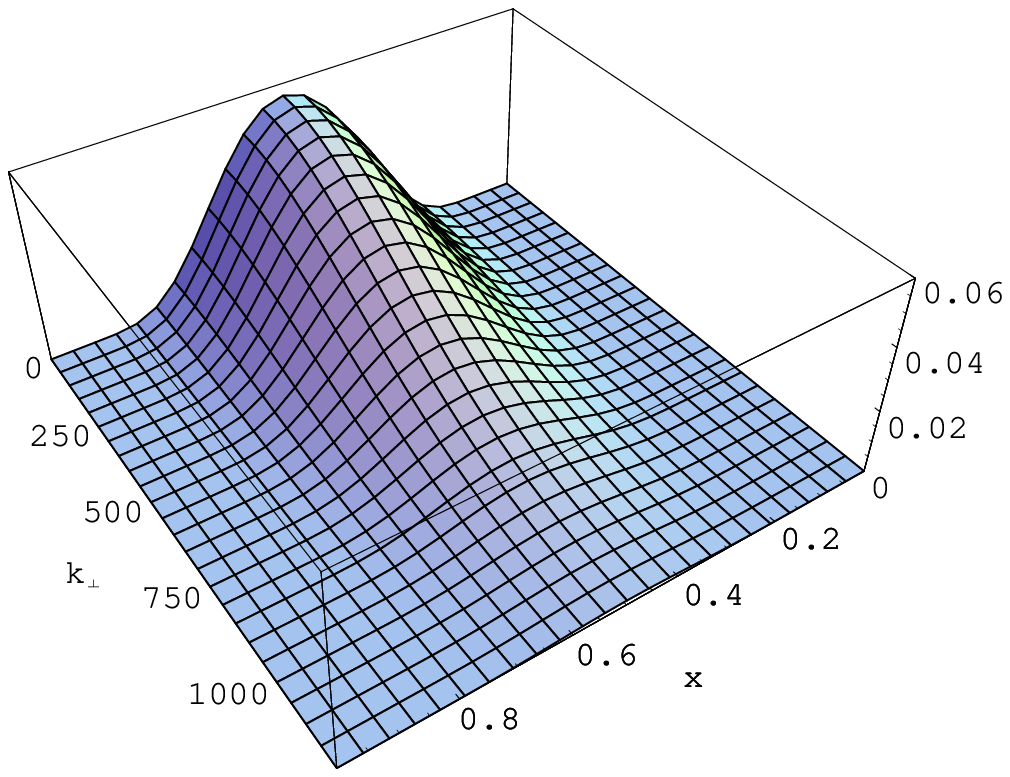}
  & \includegraphics[width=3.0in]{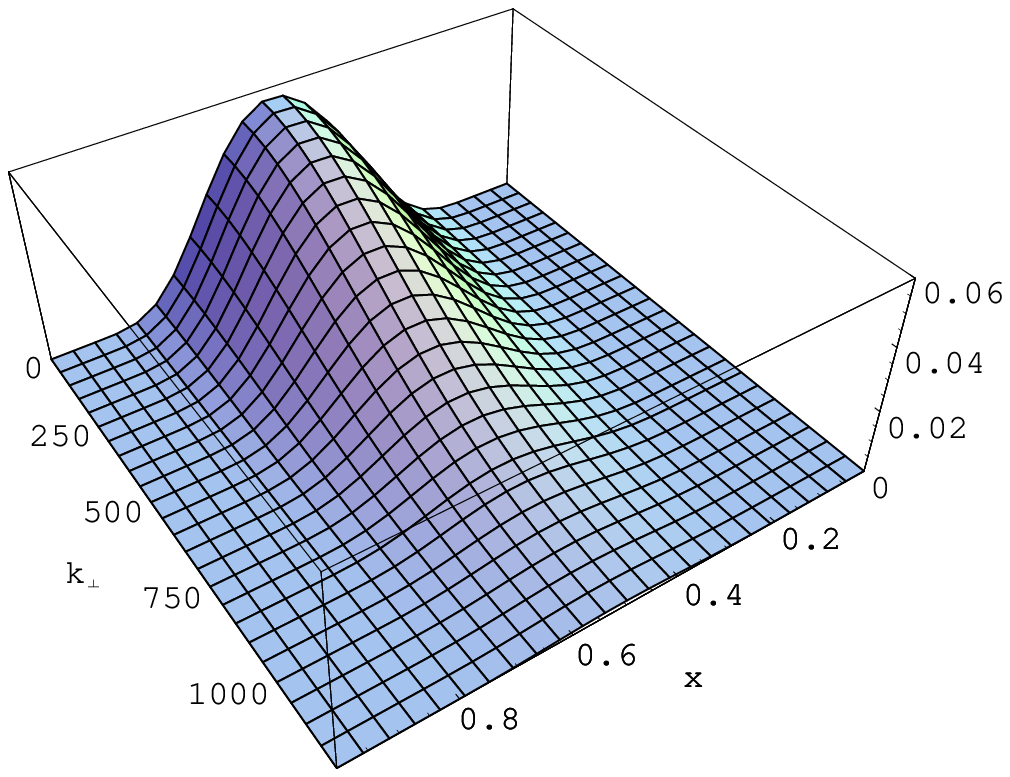}
  \end{tabular}

\caption{Wave function $\psi_{c\bar c}(x,\bfk)$ according to section
IV. We consider in this case $m_c = 1.5$ GeV and
$\kappa_1 = \kappa_2 = 894$ MeV, the value suggested by the Regge
slope for charmonium states. The left graph corresponds to Eq.~(\ref{FnOnda1})
and the right graph to Eq.~(\ref{FnOnda2}).}
\end{figure*}

\subsection{Model 2}

Model 2 has originally been developed in Ref.~\cite{VegaSchmidt2}.
It is based on the following equation of motion for the AdS modes
\begin{equation}\label{EcuationVS}
  \partial^2_\zeta  \varphi (\zeta)
- \frac{2-\beta}{\zeta} \partial_\zeta \varphi (\zeta)
  + \biggl( M^2_2 - \frac{m_{5}^{2} R^{2}}{\zeta^{2}} \biggr)  \varphi (\zeta)
  = 0 \,,
\end{equation}
where for mesons we have
\begin{equation}
 m_{5}^{2} R^{2} = (3 + L - S + \kappa^2_2 \zeta^2)
                   (L - S + \beta + \kappa^2_2 \zeta^2).
\end{equation}
Here subscript ``2'' indicates the solutions of the Model~2.
From~(\ref{EcuationVS}) we get the mass spectrum
\begin{equation}
 \label{EspectroVS}
 M^2_2 = 4 \kappa^2_2 \biggl[n + L + \biggl(2 + \frac{\beta}{2}
       - S \biggr)\biggr],
\end{equation}
where $\beta = -3$ is the value for scalar mesons and $\beta = -1$
for vector mesons \cite{VegaSchmidt2}. To have consistency with the
definition of the form factor of Eq.~(\ref{FactorFormaAdS})
the equation~(\ref{EcuationVS}) should be changed into
a Schr\"odinger type equation of
\begin{equation}
 \label{EcuationBdT2}
 \biggl[ -\frac{d^{2}}{d \zeta^{2}} +
 \frac{4 m_5^2 R^2 + \beta^2 - 6 \beta + 8}{4 \zeta^2}
 \biggr] \Phi_2(\zeta) = M^2_2 \Phi_2(\zeta) ,
\end{equation}
by means of the following transformation
\begin{equation}\label{Trans_Model2}
 \varphi (\zeta) = e^{(1 - \beta/2) \ln\zeta} \Phi_2 (\zeta)
\,.
\end{equation}
In this model we have $f(x) = 2 x$~\cite{VegaSchmidt2} and
the matching condition between the LFWF and AdS modes reads
\begin{equation}
 \label{FnOnda22}
 |\widetilde{\psi}_{q_1\bar q_2}^{(2)}(x,\zeta)|^{2} =
2 A_2^2 x^{2} (1-x)  \frac{|\Phi_2(\zeta)|^2}{2\pi\zeta} \,.
\end{equation}
Again we restrict to the ground state case --- $n = L = 0$ and
as AdS mode $\Phi_2(\zeta)$ similar to the one of Model 1:
\begin{equation}\label{Phi2_AdS}
\Phi_2(\zeta) \sim \sqrt{\zeta} e^{-\frac{1}{2} \kappa^2_2 \zeta^{2}} \,.
\end{equation}
Finally applying the Brodsky and T\'eramond
prescription, the meson momentum space LFWF including
massive quarks is
\begin{equation}\label{FnOnda2}
\hspace*{-.2cm}
 \psi_{q_1\bar q_2}^{(2)}(x,\bfk)
= \frac{4\pi A_2}{\kappa_2} \sqrt{\frac{2}{1-x}}
 \exp\biggl(-\frac{\bfk^2}{2 \kappa_2^2 x(1-x)}
-\frac{\mu_{12}^2}{2\kappa^2_2}
 \biggr) \,,
\end{equation}
where $A_2$ is the normalization constant constrained by the
probability condition~(\ref{Probability_cond}) in analogy to $A_1$.

\section{Example I: The Pion}

\subsection{Fixing the parameters}

The wave functions we consider depend on parameters
($A_i$,$m_{1,2}$,$\kappa_i$) which must be fixed.
As a first application we consider some of the fundamental
properties of the pion: leptonic and two-photon decay constants,
distribution quantities. We work in the isospin limit, supposing
that the masses of $u$ and $d$ quarks are equal to each other:
$m = m_u = m_d$. In this case we have a set of three free parameters
($A_i$,$m$,$\kappa_i$) which is the same number of parameters
considered in other models~\cite{HMS}.

The two conditions related to the decay amplitudes
for $\pi \rightarrow \mu \nu$ and $\pi^{0} \rightarrow \gamma
\gamma$~\cite{BHL} read
\begin{equation}
 \label{Condition1}
 \int\limits_0^1 dx \int \frac{d^2 \bfk}{16 \pi^{3}}
\psi_{q \overline{q}} (x,\bfk) = \frac{F_\pi}{2 \sqrt{3}},
\end{equation}
and
\begin{equation}
 \label{Condition2}
 \int\limits_0^1 dx \, \psi_{q \overline{q}} (x,\bfk = 0)
= \frac{\sqrt{3}}{F_\pi},
\end{equation}
where $F_{\pi} = f_\pi/\sqrt{2} \simeq 92.4$ MeV
is the pion leptonic decay constant.
Note, the second condition~(\ref{Condition2}) is the low-energy
theorem relating the two-photon $g_{\pi\gamma\gamma}$ and leptonic $F_\pi$
decay constants as
$g_{\pi\gamma\gamma} = 1/(4\pi^2 F_\pi) = 0.274$~GeV$^{-1}$.

On the other side, the average transverse momentum squared of
a quark in the pion $\langle  \bfk^2 \rangle_{\pi}$ is about
(300 MeV)$^2$~\cite{Metcalf}. The average transverse momentum
squared of a quark in the pion valence state is defined by
\begin{equation}
 \label{Condition3}
 \langle \bfk^2 \rangle_{q \overline{q}} =
\frac{1}{P_{q \overline{q}}}  \int\limits_0^1 dx
\int \frac{d^2\bfk}{16 \pi^{3}} \, \bfk^2 \,
|\psi_{q \overline{q}} (x,\bfk)|^{2},
\end{equation}
which must be higher than $\langle \bfk^2 \rangle_{\pi}$. For
this reason we consider a value of several hundreds of MeV for
$\sqrt{\langle \bfk^2 \rangle_{q \overline{q}}}$. This can be
used as a third restriction.
When fixing the parameters we consider two cases for each wave
function (\ref{FnOnda1}) and (\ref{FnOnda2}),  current and
constituent quark masses. The values used are 4 MeV for current masses
and 330 MeV for constituent masses.

Since quarks masses are introduced in advance, the remaining
parameters $A_i$ and $\kappa_1$ or $\kappa_2$ can be fixed using
(\ref{Condition1}) and (\ref{Condition2}) with the value of
$F_{\pi} = 92.4$ MeV. Then with the fixed parameters
$A_i, m, \kappa_i$ we predict
$\sqrt{\langle \bfk^2 \rangle_{q\overline{q}}}$
and the probability $P_{q \overline{q}}$. Table I gives the values
for $A_{1,2}$ and $\kappa_{1,2}$ including the predictions for
$\sqrt{\langle \bfk^2 \rangle_{q\overline{q}}}$ and $P_{q \overline{q}}$.
One can see that our results for
$\sqrt{\langle \bfk^2 \rangle_{q\overline{q}}}$ and $P_{q \overline{q}}$
are in agreement with the predictions of Ref.~\cite{HMS}:
$\sqrt{\langle \bfk^2 \rangle_{q\overline{q}}} \simeq $ 356 MeV and
$P_{q \overline{q}} \simeq $ 0.296.

\begin{table*}
\begin{center}
\caption{Parameters defining LFWF given by Eqs.~(\ref{FnOnda1})
and (\ref{FnOnda2}) and predictions for $\sqrt{\langle
\bfk^2 \rangle_{q \overline{q}}}$ and $P_{q \overline{q}}$.}

%\vspace*{.2cm}
\begin{tabular}{| c | c | c | c | c | c | c |}
  \hline
Model & $\psi (x,\bfk)$ & $m$ (MeV) & $A$ & $\kappa$ (MeV) &
$\sqrt{\langle \bfk^2 \rangle_{q \overline{q}}}$ (MeV) &
$P_{q \overline{q}}$  \\
  \hline
  & $\psi_{1c} $ & 4   & 0.452  & 951.043 & 388.319 & 0.204 \\
1 &              &     &        &     &     &      \\
  & $\psi_{1cs}$ & 330 & 0.924   & 787.43 & 356.478 & 0.279 \\
\hline
  & $\psi_{2c} $ & 4   & 0.486   & 921.407 & 376.222 & 0.236 \\
2 &              &     &        &     &     &      \\
  & $\psi_{2cs}$ & 330 & 0.965 & 781.218 & 353.877 & 0.299 \\
  \hline
\end{tabular}
\end{center}
\end{table*}

The parameters $\kappa_{1,2}$ define the holographic model
considered in Refs.~\cite{BdT4,VegaSchmidt2} and both are related
to the Regge slope. Thus in principle both quantities could be fixed
by spectral data. Unfortunately the pion mass is an exception since
it falls outside the Regge trajectories. Therefore $\kappa_{1,2}$
have been usually fixed by using form factors~\cite{BdT4, VegaSchmidt2}.
The values obtained in the present work differ somewhat from those
values, which is understandable since the $\kappa_1$ and $\kappa_2$
found previously were obtained using~(\ref{FactorFormaAdS}), the
form factor in AdS, which when compared with the light front
expression gave~(\ref{Densidad}). Nevertheless, the wave
functions~(\ref{FnOnda1}) and~(\ref{FnOnda2}) correspond to the case
with only two quarks, and we therefore should expect a small change
in the $\kappa_{1,2}$ values.

\subsection{Pion Distribution Amplitude}

The meson distribution amplitude is calculated using~\cite{Lepage1}
\begin{equation}\label{phi_xq}
 \phi (x,q) = \int^{q^{2}} \frac{d^{2} \bfk}{16 \pi^{3}}
\psi_{\rm val} (x, \bfk).
\end{equation}
We remind that the pion $|\psi\ra$ can be expanded into Fock states
$|\psi\ra = a_{1} |q \overline{q} \rangle + a_{2} |q \overline{q} g \rangle +
a_{3} |q \overline{q} g g \rangle + ... $. For large values of $q^{2}$
the dominant term is the first one and since our wave functions were
obtained considering (\ref{Densidad2Partones}), which corresponds to
the $q\bar q$ configuration, we can calculate
$\phi (x) \equiv \phi (x, Q \rightarrow \infty)$.

Using (\ref{FnOnda1}) and (\ref{FnOnda2}) we get
\begin{equation}
 \label{Phi1}
 \phi_{1} (x) = \frac{A_1\kappa_1}{2 \pi} \sqrt{x(1-x)}
\exp\Big(- \frac{m^2}{2 \kappa^2_1 x(1-x)}\Big) \,,
\end{equation}
and
\begin{equation}
 \label{Phi2}
 \phi_{2} (x) = \frac{A_2\kappa_2}{2 \pi} x \sqrt{2(1-x)}
\exp\Big(- \frac{m^2}{2 \kappa^2_1 x(1-x)}\Big) \,.
\end{equation}
In Fig.4 both expressions are compared for current (c) and
constituent (cs) quark masses to the prediction of PQCD using
$\phi (x, Q \rightarrow \infty) = \sqrt{3} F_{\pi} x (1-x)$~\cite{Lepage2}.
\begin{figure}
  \begin{tabular}{cc}
    \includegraphics[width=3.0in]{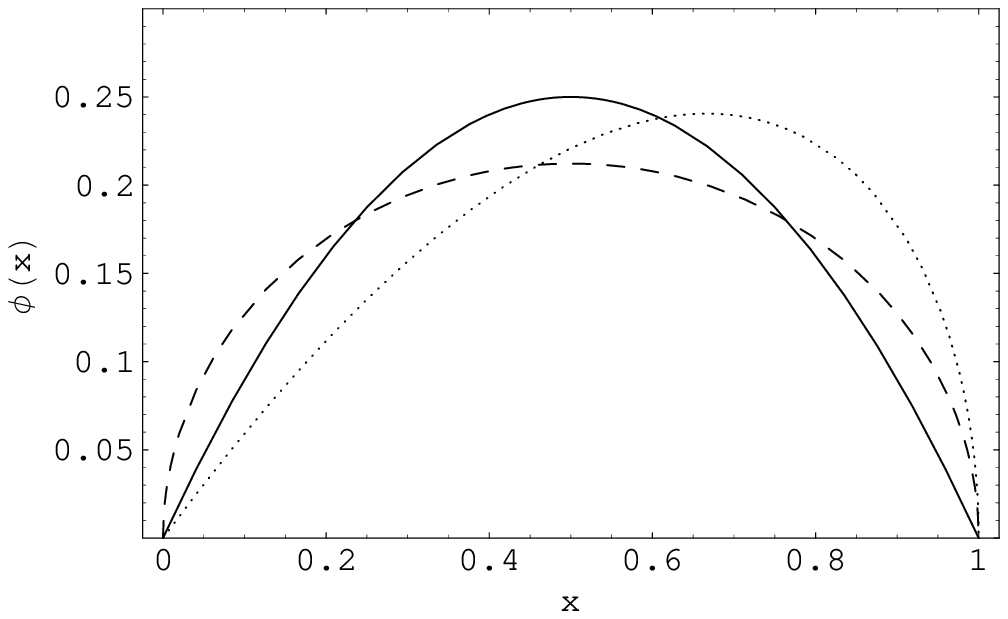}
  \end{tabular}
  \begin{tabular}{cc}
    \includegraphics[width=3.0in]{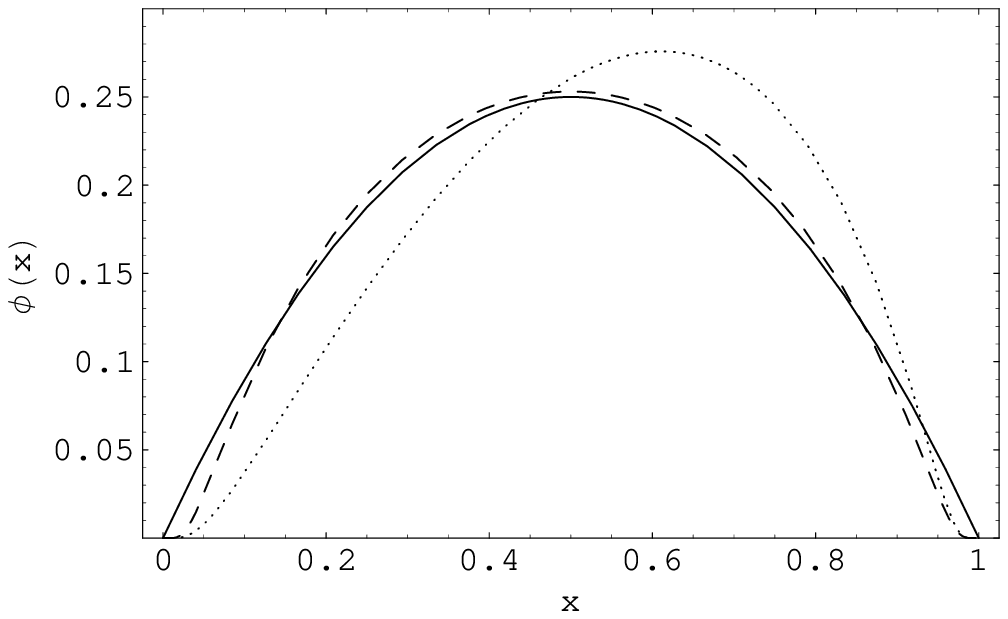}
  \end{tabular}
\caption{Pion distribution amplitudes using holographic LFWF.
Solid lines correspond to the prediction of PQCD,
dashed lines correspond to LFWF~(\ref{Phi1}), and the dotted
ones to LFWF~(\ref{Phi2}) for $m = 4$ MeV (upper panel)
and for $m = 330$ MeV (lower panel).}
\end{figure}
Fig.4 shows that increasing quark masses reduces the differences
between the two variants of LFWFs.
Knowing the distribution amplitudes, it is possible to calculate
the moments. Taking $\xi = 1 - 2 x$ we have
\eq\label{xiN}
 \langle \xi^{N} \rangle = \frac{\int\limits_{-1}^{1} d\xi \xi^N
\phi(\xi)}{\int\limits_{-1}^{1} d\xi \phi (\xi)} \,.
\en
Table II contains a summary of the moments up to $\langle \xi^{4} \rangle$.

\begin{table}
\begin{center}
\caption{First moments of the distribution functions
$\langle \xi^{N} \rangle$ calculated using $\phi_{\rm PQCD}$
and $\phi$, given explicitly by ~(\ref{Phi1}) and~(\ref{Phi2}),
for $m = 4$ MeV and $m = 330$ MeV. For $\phi_{2cs\ast}$
we take $m=300$ MeV, which shows that odd moments are reduced
when the quark mass quarks increases.}
\begin{tabular}{ c c c | c c c | c c c | c c c | c c c | c c c }
  \hline
  & $\phi$ & & & $\langle \xi^{0} \rangle$
& & & $\langle \xi^{1} \rangle$
& & & $\langle \xi^{2} \rangle$
& & & $\langle \xi^{3} \rangle$
& & & $\langle \xi^{4} \rangle$ & \\
  \hline
  & $\phi_{\rm PQCD}$ & & & 1 & & & 0 & & & 0.2 & & & 0 & & & 0.086 & \\
  & $\phi_{1c}$ & & & 1 & & & 0 & & & 0.250 & & & 0 & & & 0.125 & \\
  & $\phi_{2c}$ & & & 1 & & & 0.143 & & & 0.238 & & & 0.073 & & & 0.116 & \\
  & $\phi_{1cs}$ & & & 1 & & & 0 & & & 0.186 & & & 0 & & & 0.073 & \\
  & $\phi_{2cs}$ & & & 1 & & & 0.102 & & & 0.179 & & & 0.040 & & & 0.068 & \\
  & $\phi_{2cs\ast}$ & & & 1 & & & 0.106 & & & 0.187 & & & 0.044 & & & 0.073&\\
  \hline
\end{tabular}
\end{center}
\end{table}

\subsection{Parton distributions}

If the LFWF has the form
\begin{equation}
 \label{FnOndaTipo}
 \psi_{q\bar q}(x,\bfk) = \eta (x)
\exp\Big(\frac{\bfk^2}{2 \lambda^2 x (1 - x)} \Big) \,,
\end{equation}
then the parton distribution is given by~\cite{Radyushkin}
\begin{equation}
 \label{DensidadPartones}
 f(x) = \frac{x(1-x) \lambda^{2}}{16 \pi^{2}} \eta^{2} (x) \,.
\end{equation}

The LFWFs obtained from Models 1 and 2 have the form considered in
(\ref{FnOndaTipo}) and then the two-body contribution to the parton
distributions can be calculated in a direct way. In Fig.5 we display
the product $x f(x)$ for both models again using current and
constituent quark masses in the LFWF. We use the same parameters
as in Table~I.

In principle, contributions from higher Fock states should be added
because they are not necessarily small. In fact, in the pion
case that we are discussing here, the valence state component is around
25 percent as can be seen in Table I or for example in Refs.~\cite{HMS,BHL}.

\begin{figure*}
  \begin{tabular}{cc cc}
    \includegraphics[width=3.0in]{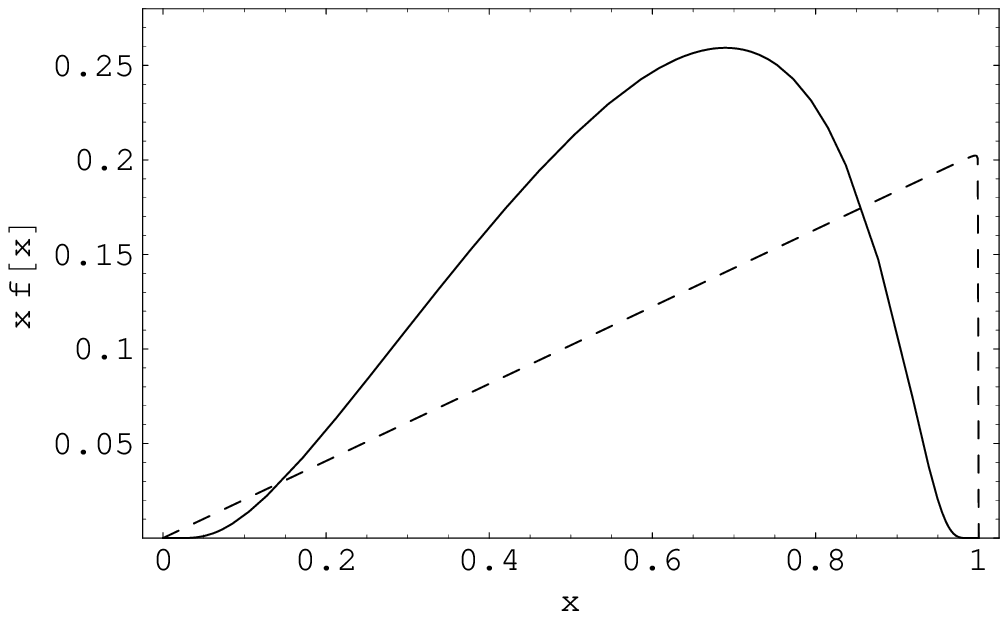}
  & \includegraphics[width=3.0in]{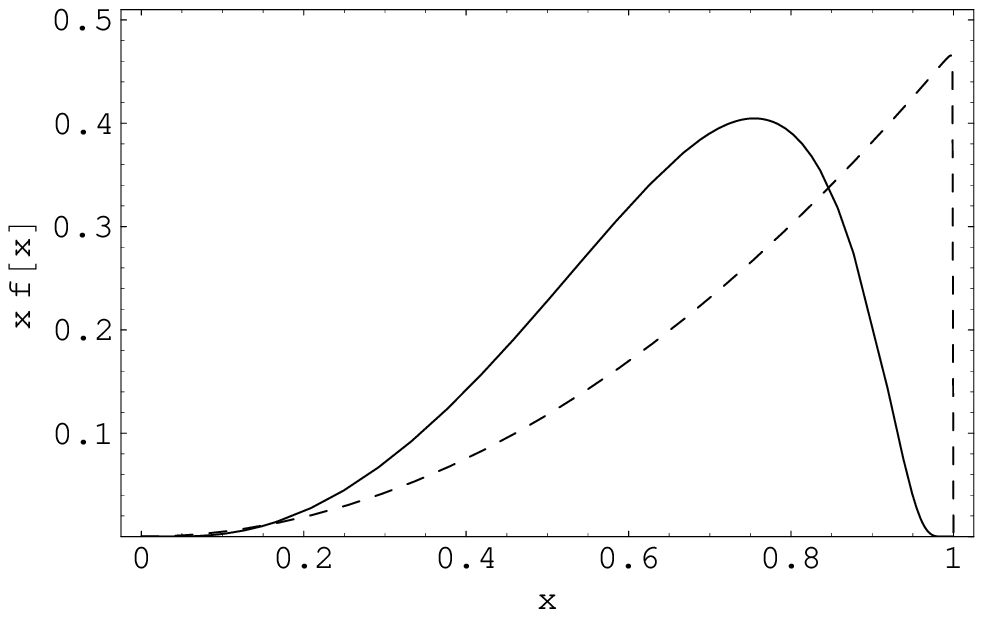}
  \end{tabular}
\caption{Valence parton distribution $x f(x)$
according to the LFWF considered in this work. The right graph
corresponds to model 1 and the left graph is for model 2. In both
cases the dashed line corresponds to the case with current masses,
while the solid line correspond to the constituent mass case. The
parameters involved are the same as displayed in Table I.}
\end{figure*}

\section{Example II: Decay Constants}

Now we are in the position to calculate leptonic couplings of
pseudoscalar $(f_P)$ and vector $(f_V)$ mesons which are given
in our approach by
\begin{equation}
 \label{ConsDecay}
f_{P} = f_{V} = 2 \sqrt{6} \int^{1}_{0} dx
\int \frac{d^2 \bfk}{16 \pi^{3}} \psi_{q \overline{q}} (x,\bfk) \,.
\end{equation}
We use experimental values for the decay constants and the
probability condition
\begin{equation}
\label{Normalization}
P_{q\bar q} = \int\limits_0^1 dx \int \frac{d^{2} \bfk}{16 \pi^{3}}
|\psi_{q \overline{q}} (x,\bfk)|^{2} \leq 1 ,
\end{equation}
where the equality holds for the case when the valence part
dominates. This procedure allows to fix the parameters $\kappa_{1,2}$
and the normalization constants $A_{1,2}$.

Holographic models usually give a relation
between $\kappa_{1,2}$ and the Regge slope fixed by
spectroscopic data. Thus the only free
parameter $A_{1,2}$ can be fixed by the normalization condition.
As an example we consider the decay constant for kaons and $J/\psi$
assuming the valence contribution to be dominant, i.e we
use~(\ref{Normalization}) with $P_{q\bar q}=1$. The quark masses used are
\begin{eqnarray}
m_{u} = m_{d} & = & 330 \ {\rm MeV}\,, \nonumber \\
m_{s} & = & 500 \ {\rm MeV}\,, \nonumber \\
m_{c} & = & 1500 \ {\rm MeV}\,. \nonumber
\end{eqnarray}

As already mentioned, the parameters $\kappa_{1,2}$ can be
fixed by using Regge slope data~\cite{Iachello,Gershtein}:
for kaon we take $\kappa_1 = \kappa_2 = 524$ MeV~\cite{Iachello},
while for $J/\psi$ we use $\kappa_1 = \kappa_2 = 894$~MeV.

Now we can calculate the decay constants of $K$ and $J/\psi$ mesons.
In Model 1 we obtain:
$f_K = 156.01$ MeV and
$f_{J/\psi} = 226.68$ MeV. Our predictions in Model 2 are:
$f_K = 156.35$ MeV and
$f_{J/\psi} = 224.97$ MeV. Our results for the $\pi$ and $K$ meson
decay constants in both models are close to the experimental values
of 155.5 and 277.6 MeV, respectively.

Further applications of the approach considered here to the mass
spectrum and decay constants of light and heavy hadrons will
be considered in Ref.~\cite{our_paper2}.

\section{Conclusions}

We have considered two kinds of wave functions for mesons in the
light-front formalism obtained by the AdS/CFG correspondence
with two soft wall holographic models.
By identifying in the momentum space wave function the kinetic energy in
the massless case we could introduce the quark mass dependence
as suggested by Brodsky and Teramond~\cite{BdT5}. Both wave functions
have a different $x$ dependence, which is less pronounced when the
quark masses are increased as can be seen in Figs.1-3.

If we restrict ourselves to pions, there is an asymmetry in
(\ref{FnOnda2}) which does not appear in (\ref{FnOnda1}), although
this is reduced when using constituent masses. One motivation to use
(\ref{FnOnda2}) is that it was obtained from a more general
holographic soft wall model than the one that considers a
quadratic dilaton.

But when other mesons are considered, it is important
to note that the parameters $\kappa_{1,2}$ used in the
holographic models can be fixed by spectroscopic data, since
these parameters are related to the Regge trajectory. Taking quark
masses as initial input only one parameter remains
(the normalization constant $A_{1,2}$) which can be fixed by
the normalization condition.

Due to the importance of the hadronic wave function in QCD the
versions considered in this work represent a clear example of the
usefulness of the AdS/CFT ideas in QCD applications.
These wave functions can be used as initial ansatz in variational
treatments or as a first step in order to diagonalize the light
front QCD Hamiltonian.

Another aspect that was not considered here is related to the fact
that the AdS modes dual to mesons have a dependence
on $n$ and $L$~\cite{HMS,VegaSchmidt2}, the radial and
angular quantum numbers
respectively. Thus in principle it should be possible to obtain
LFWFs for radial and angular excitations also.
The Gauge/Gravity dualities offer an interesting opportunity
to consider different meson excitations and in future work we plan
to see whether these models reproduce the corresponding data
in these cases.

\begin{acknowledgments}

The A.V. work was partially supported by DGIP from Universidad
T\'ecnica Federico Santa Mar\'ia. A.V. is grateful for the
hospitality of the Institut f\"ur Theoretische Physik of
Universit\"at T\"ubingen, where this work started.
This work was supported by the DFG under Contract No. FA67/31-1,
No. FA67/31-2, and No. GRK683. This research is also part of the
European Community-Research Infrastructure Integrating Activity
``Study of Strongly Interacting Matter'' (HadronPhysics2,
Grant Agreement No. 227431) and of the President grant of Russia
``Scientific Schools''  No. 871.2008.2.

\end{acknowledgments}


\begin{thebibliography}{99}

\bibitem{BetheSalpeter}
  E.~E.~Salpeter and H.~A.~Bethe,
  %``A Relativistic equation for bound state problems,''
  Phys.\ Rev.\  {\bf 84}, 1232 (1951).
  %%CITATION = PHRVA,84,1232;%%

\bibitem{PauliBrodsky}
  H.~C.~Pauli and S.~J.~Brodsky,
  %``Solving Field Theory In One Space One Time Dimension,''
  Phys.\ Rev.\  D {\bf 32}, 1993 (1985);
  %%CITATION = PHRVA,D32,1993;%%
  %``Discretized Light Cone Quantization: Solution To A Field Theory In One
  %Space One Time Dimensions,''
  Phys.\ Rev.\  D {\bf 32}, 2001 (1985).
  %%CITATION = PHRVA,D32,2001;%%

\bibitem{BJS}
  S.~J.~Brodsky, C.~R.~Ji and M.~Sawicki,
  %``Evolution Equation And Relativistic Bound State Wave Functions For Scalar
  %Field Models In Four And Six-Dimensions,''
  Phys.\ Rev.\  D {\bf 32}, 1530 (1985).
  %%CITATION = PHRVA,D32,1530;%%

\bibitem{JK}
  O.~C.~Jacob and L.~S.~Kisslinger,
  %``The Pion form-factor in a light cone representation,''
  Phys.\ Lett.\  B {\bf 243}, 323 (1990).
  %%CITATION = PHLTA,B243,323;%%

\bibitem{PolStrass1}
  J.~Polchinski and M.~J.~Strassler,
  %``Hard scattering and gauge / string duality,''
  Phys.\ Rev.\ Lett.\  {\bf 88}, 031601 (2002)
  [arXiv:hep-th/0109174].
  %%CITATION = PRLTA,88,031601;%%

\bibitem{Janik}
  R.~A.~Janik and R.~B.~Peschanski,
  %``High energy scattering and the AdS/CFT correspondence,''
  Nucl.\ Phys.\  B {\bf 565}, 193 (2000)
  [arXiv:hep-th/9907177].
  %%CITATION = NUPHA,B565,193;%%

\bibitem{BdT1}
  S.~J.~Brodsky and G.~F.~de Teramond,
  %``Light-front hadron dynamics and AdS/CFT correspondence,''
  Phys.\ Lett.\  B {\bf 582}, 211 (2004)
  [arXiv:hep-th/0310227].
  %%CITATION = PHLTA,B582,211;%%

\bibitem{Levin}
  E.~Levin, J.~Miller, B.~Z.~Kopeliovich and I.~Schmidt,
  %``Glauber - Gribov approach for DIS on nuclei in N=4 SYM,''
  JHEP {\bf 0902}, 048 (2009)
  [arXiv:0811.3586 [hep-ph]].
  %%CITATION = JHEPA,0902,048;%%

\bibitem{BdT2}
  G.~F.~de Teramond and S.~J.~Brodsky,
  %``The hadronic spectrum of a holographic dual of QCD,''
  Phys.\ Rev.\ Lett.\  {\bf 94}, 201601 (2005)
  [arXiv:hep-th/0501022].
  %%CITATION = PRLTA,94,201601;%%

\bibitem{KKSS}
  A.~Karch, E.~Katz, D.~T.~Son and M.~A.~Stephanov,
  %``Linear Confinement and AdS/QCD,''
  Phys.\ Rev.\  D {\bf 74}, 015005 (2006)
  [arXiv:hep-ph/0602229].
  %%CITATION = PHRVA,D74,015005;%%

\bibitem{Forkel}
  H.~Forkel, M.~Beyer and T.~Frederico,
  %``Linear square-mass trajectories of radially and
  %orbitally excited hadrons in holographic QCD,''
  JHEP {\bf 0707}, 077 (2007)
  [arXiv:0705.1857 [hep-ph]].
  %%CITATION = JHEPA,0707,077;%%

\bibitem{VegaSchmidt1}
  A.~Vega and I.~Schmidt,
  %``Scalar hadrons in $AdS_{5} \times S^{5}$,''
  Phys.\ Rev.\  D {\bf 78}, 017703 (2008)
  [arXiv:0806.2267 [hep-ph]].
  %%CITATION = PHRVA,D78,017703;%%

\bibitem{VegaSchmidt2}
  A.~Vega and I.~Schmidt,
  %``Hadrons in AdS(5) x S**5,''
  Phys.\ Rev.\  D {\bf 79}, 055003 (2009)
  [arXiv:0811.4638 [hep-ph]].
  %%CITATION = PHRVA,D79,055003;%%

\bibitem{DaRol}
  L.~Da Rold and A.~Pomarol,
  %``Chiral symmetry breaking from five dimensional spaces,''
  Nucl.\ Phys.\  B {\bf 721}, 79 (2005)
  [arXiv:hep-ph/0501218].
  %%CITATION = NUPHA,B721,79;%%

\bibitem{EKSS}
  J.~Erlich, E.~Katz, D.~T.~Son and M.~A.~Stephanov,
  %``QCD and a Holographic Model of Hadrons,''
  Phys.\ Rev.\ Lett.\  {\bf 95}, 261602 (2005)
  [arXiv:hep-ph/0501128].
  %%CITATION = PRLTA,95,261602;%%

\bibitem{Colangelo}
  P.~Colangelo, F.~De Fazio, F.~Giannuzzi, F.~Jugeau and S.~Nicotri,
  %``Light scalar mesons in the soft-wall model of AdS/QCD,''
  Phys.\ Rev.\  D {\bf 78}, 055009 (2008) \\{}
  [arXiv:0807.1054 [hep-ph]].
  %%CITATION = PHRVA,D78,055009;%%

\bibitem{Boschi}
  H.~Boschi-Filho, N.~R.~F.~Braga and C.~N.~Ferreira,
  %``Static strings in Randall-Sundrum scenarios and the quark anti-quark
  %potential,''
  Phys.\ Rev.\  D {\bf 73}, 106006 (2006)
  [Erratum-ibid.\  D {\bf 74}, 089903 (2006)]
  [arXiv:hep-th/0512295].
  %%CITATION = PHRVA,D73,106006;%%

\bibitem{Andreev}
  O.~Andreev and V.~I.~Zakharov,
  %``Heavy-quark potentials and AdS/QCD,''
  Phys.\ Rev.\  D {\bf 74}, 025023 (2006)
  [arXiv:hep-ph/0604204].
  %%CITATION = PHRVA,D74,025023;%%

\bibitem{Jugeau}
  F.~Jugeau,
  %``Hadrons potentials within the gauge/string correspondence,''
  arXiv:0812.4903 [hep-ph].
  %%CITATION = ARXIV:0812.4903;%%

\bibitem{Hambye}
  T.~Hambye, B.~Hassanain, J.~March-Russell and \\{} M.~Schvellinger,
  %``On the Delta(I) = 1/2 rule in holographic QCD,''
  Phys.\ Rev.\  D {\bf 74}, 026003 (2006) \\{}
  [arXiv:hep-ph/0512089];
  %%CITATION = PHRVA,D74,026003;%%
  %``Four-point functions and kaon decays in a minimal AdS/QCD model,''
  Phys.\ Rev.\  D {\bf 76}, 125017 (2007)
  [arXiv:hep-ph/0612010].
  %%CITATION = PHRVA,D76,125017;%%

\bibitem{BdT3}
  S.~J.~Brodsky and G.~F.~de Teramond,
  %``Hadronic spectra and light-front wavefunctions in holographic QCD,''
  Phys.\ Rev.\ Lett.\  {\bf 96}, 201601 (2006)
  [arXiv:hep-ph/0602252].
  %%CITATION = PRLTA,96,201601;%%

\bibitem{BdT4}
  S.~J.~Brodsky and G.~F.~de Teramond,
  %``Light-Front Dynamics and AdS/QCD Correspondence: The Pion Form Factor in
  %the Space- and Time-Like Regions,''
  Phys.\ Rev.\  D {\bf 77}, 056007 (2008)
  [arXiv:0707.3859 [hep-ph]].
  %%CITATION = PHRVA,D77,056007;%%

\bibitem{BPP}
  S.~J.~Brodsky, H.~C.~Pauli and S.~S.~Pinsky,
  %``Quantum Chromodynamics and Other Field Theories on the Light Cone,''
  Phys.\ Rept.\  {\bf 301}, 299 (1998)
  [arXiv:hep-ph/9705477].
  %%CITATION = PRPLC,301,299;%%

\bibitem{Grigoryan}
  H.~R.~Grigoryan and A.~V.~Radyushkin,
  %``Structure of Vector Mesons in Holographic Model with Linear Confinement,''
  Phys.\ Rev.\  D {\bf 76}, 095007 (2007)
  [arXiv:0706.1543 [hep-ph]].
  %%CITATION = PHRVA,D76,095007;%%

\bibitem{BdT6}
  S.~J.~Brodsky and G.~F.~de Teramond,
  %``Light-Front Holography and Novel Effects in QCD,''
  arXiv:0812.3192 [hep-ph].
  %%CITATION = ARXIV:0812.3192;%%

\bibitem{BdT5} See, e.g.,
  S.~J.~Brodsky and G.~F.~de Teramond, \\{}
  %``AdS/CFT and Light-Front QCD,''
  arXiv:0802.0514 [hep-ph].
  %%CITATION = ARXIV:0802.0514;%%

\bibitem{HMS}
  T.~Huang, B.~Q.~Ma and Q.~X.~Shen,
  %``Analysis of the pion wave function in light cone formalism,''
  Phys.\ Rev.\  D {\bf 49}, 1490 (1994)
  [arXiv:hep-ph/9402285].
  %%CITATION = PHRVA,D49,1490;%%

\bibitem{BHL}
  S.~J.~Brodsky, T.~Huang and G.~P.~Lepage,
  %``Hadronic Wave Functions And High Momentum Transfer Interactions In Quantum
  %Chromodynamics,''
  Proceedings of the Banff Summer Institute ``Particles and Fields 2'',
  Banff, Alberta, 1981, edited by A.~Z.~Capri and A.~N.~Kamal
  (Plenum, New York, 1983), p. 143;
  P.~Lepage, S.~J.~Brodsky, T.~Huang and P.~B.~Mackenzie, ibid., p. 83;
  T.~Huang,
  %``Hadron Wave Functions And Structure Functions In QCD,''
  AIP Conf.\ Proc.\  {\bf 68}, 1000 (1981).

\bibitem{Iachello}
  F.~Iachello, N.~C.~Mukhopadhyay and L.~Zhang,
  %``Spectrum Generating Algebra For String Like Mesons. 1. Mass Formula For Q
  %Anti-Q Mesons,''
  Phys.\ Rev.\  D {\bf 44}, 898 (1991).
  %%CITATION = PHRVA,D44,898;%%

\bibitem{Gershtein}
  S.~S.~Gershtein, A.~K.~Likhoded and A.~V.~Luchinsky,
  %``Systematics of heavy quarkonia from Regge trajectories on (n,M**2) and
  %(M**2,J) planes,''
  Phys.\ Rev.\  D {\bf 74}, 016002 (2006)
  [arXiv:hep-ph/0602048].
  %%CITATION = PHRVA,D74,016002;%%

\bibitem{Metcalf} See, e.g.,
  W.~J.~Metcalf {\it et al.},
  %``The Magnitude Of Parton Intrinsic Transverse Momentum,''
  Phys.\ Lett.\  B {\bf 91}, 275 (1980).
  %%CITATION = PHLTA,B91,275;%%

\bibitem{Lepage1}
  G.~P.~Lepage and S.~J.~Brodsky,
  %``Exclusive Processes In Perturbative Quantum Chromodynamics,''
  Phys.\ Rev.\  D {\bf 22}, 2157 (1980).
  %%CITATION = PHRVA,D22,2157;%%

\bibitem{Lepage2} G
  G.~P.~Lepage and S.~J.~Brodsky,
  %``Exclusive Processes In Quantum Chromodynamics: Evolution Equations For
  %Hadronic Wave Functions And The Form-Factors Of Mesons,''
  Phys.\ Lett.\  B {\bf 87}, 359 (1979).
  %%CITATION = PHLTA,B87,359;%%

\bibitem{Radyushkin}
  A.~V.~Radyushkin,
  %``Nonforward parton densities and soft mechanism for form factors and
  %wide-angle Compton scattering in {QCD},''
  Phys.\ Rev.\  D {\bf 58}, 114008 (1998)
  [arXiv:hep-ph/9803316].
  %%CITATION = PHRVA,D58,114008;%%

\bibitem{our_paper2}T.~Branz, T.~Gutsche, S.~Kovalenko,
V.~E.~Lyubovitskij, I.~Schmidt, A.~Vega, in preparation.

\end{thebibliography}
\end{document}